\documentstyle[12pt]{article}
\def\CRAS{C.~R.~Acad.~Sc.~Paris}
\def \ie {i.~e.~}
\def \D {\hbox{d}}
\def \Log {\mathop{\rm Log}\nolimits}
\def \mod#1{\vert #1 \vert}
\def \bfE {{\bf E}}
\def \bfK {{\bf K}}

\def \bfR {{\bf R}}
\def \bfp {{\bf p}}
\def \bfu {{\bf u}}
\def \bfv {{\bf v}}
\def \bfx {{\bf x}}

\begin{document}

\pagestyle{plain} 

\vskip 0.3 truecm

\begin{center}
 {\bf Perturbative methods for the Painlev\'e test}
\end{center}

\vskip 0.5 truecm

{\bf R.~Conte}

\medskip
Service de physique de l'\'etat condens\'e,
CEA Saclay,
\hfill \break \indent
F--91191 Gif-sur-Yvette Cedex,
France
\medskip

\baselineskip=12truept
\vskip 0.8 truecm

{\it Abstract}.
There exist many situations where an ordinary differential equation admits a
movable critical singularity which the test of Kowalevski and Gambier fails to
detect.
Some possible reasons are~:
existence of negative Fuchs indices,
insufficient number of Fuchs indices,
multiple family,
absence of an algebraic leading order.
Mainly giving examples, 
we present the methods which answer all these questions.
They are all based on the theorem of perturbations of Poincar\'e
and computerizable.

\vfill
Nonlinear dynamics~: integrability and chaos, Tiruchirapalli, 12--16 Feb 1998,

ed.~S.~Daniel.
\hfill 
29~July~1998
\hskip 1.0 truecm
S98/048
\hskip 1.0 truecm
solv-int/9812007
\eject

\tableofcontents
\vfill \eject

The ``usual Painlev\'e test'' of Kowalevski and Gambier \cite{GambierThese},
also called method of pole-like expansions,
establishes necessary conditions for the {\it Painlev\'e property} (PP)
(defined as the absence of movable critical singularities 
in the general solution of an ordinary differential equation).
It does so by checking the existence of all converging Laurent series 
with a finite principal part susceptible to represent 
either the general solution or a particular solution.

Whenever there exists a movable multivaluedness in the general solution
and the test of Kowalevski and Gambier fails to detect it,
there exists a plethora of other methods able to perform this detection.
The purpose of these notes is to explain these algorithmic methods
not in full theoretical detail but on examples selected for their simplicity.

More details can be found in the lecture notes of a Carg\`ese school
\cite{Cargese96Conte}.

\section{Examples to be settled}
\indent

The following three ODEs possess a general solution with movable logarithms
undetected by the ``usual Painlev\'e test'' of Kowalevski and Gambier
\begin{eqnarray}
& &
u'' + 4 u u' + 2 u^3=0,
\label{eqOrder2WithLog}
\\
& &
u'''' + 3 u u'' - 4 u'^2=0,
\label{eqBureauOrder4}
\\
& &
u''' + u u'' - 2 u'^2=0.
\label{eqChazy1918Order3}
\label{eqNoDomOrder3}
\end{eqnarray}

The first equation (\ref{eqOrder2WithLog})
has a single {\it family of movable singularities}
 $u \sim u_0 \chi^p, \chi=x-x_0$
(one should avoid the term {\it branch} which induces a confusion with
branching i.e.~multivaluedness)
\begin{eqnarray}
& &
p=-1,\
E_0= u_0 (u_0-1)^2=0,\
\hbox{indices } (-1,0),\
u=\chi^{-1},
\end{eqnarray}
with the puzzling fact that $u_0$ should be at the same time
equal to $1$ according to the equation $E_0=0$,
and arbitrary according to the Fuchs index $0$.
The Laurent series is here reduced to its first term and only depends on one
arbitrary constant.
The movable logarithms, initially found by the $\alpha-$method
(\cite{PaiBSMF} \S 13, p.~221),
are exhibited in section \ref{sectionMethodPerturbativeFuchsian}.

The second equation (\ref{eqBureauOrder4}),
isolated by Bureau (\cite{BureauMII} p.~79),
possesses two families
\begin{eqnarray}
& &
p=-2, u_0=-60, \hbox{ indices } (-3,-2,-1,20),\
u=-60 \chi^{-2} + u_{20} \chi^{18} + \dots
\\
& &
p=-3, u_0 \hbox{ arbitrary}, \hbox{ indices } (-1,0),\
u=u_0 \chi^{-3} - 60 \chi^{-2}.
\label{eqBureau4p3}
\end{eqnarray}
The first family has enough indices but not enough positive ones,
while the second one has not enough indices,
therefore none of the two families can represent the general solution.
The movable logarithms are found by two methods,
in sections \ref{sectionMethodPerturbativeFuchsian}
and         \ref{sectionMethodPerturbativeNonFuchsian}.

The third equation has no power-law leading behaviour.
Chazy \cite{Chazy1912c,Chazy1918}
had to establish a special theorem, using divergent series,
to exhibit the movable logarithms.
The failure appears in section \ref{sectionMiscellaneous perturbations}.

A feature common of the method of Kowalevski when applied
to these three ODEs is the impossibility to 
represent the {\it general} solution by some
Laurent series with a finite principal part.
Accordingly, the possible presence of 
multivaluedness in the missing part of the general solution
cannot be tested by that method.

The common principle to all the methods performing such a detection
is to perturb a particular solution into the general solution.
Let us first recall the relevant theorems.

\section{Basic theorems}
\indent

Boldface letters denote multicomponent quantities.

{\it Theorem of perturbations}
 (Poincar\'e, {\it M\'ecanique c\'eleste} \cite{Poincare}).
Consider an ODE of order $N$, of degree one in the highest derivative,
depending on a small complex parameter $\varepsilon$,
defined in the canonical form
\begin{equation}
\label{eqLemma}
{\D \bfu \over \D x}=\bfK[x,\bfu,\varepsilon],\
   x \in {\cal C},\ \bfu \in {\cal C}^N,\ \varepsilon \in {\cal C}.
\end{equation}
Let $(x_0,\bfu_0,0)$ be a point in
${\cal C} \times {\cal C}^N \times {\cal C}$
and $D$ be a domain containing $(x_0,\bfu_0,0).$
If $\bfK$ is holomorphic in $D$,
\begin{itemize}
\item{} there {\it exists} a solution $\bfu(x,\varepsilon)$
        satisfying the initial condition $\bfu(x_0,0)=\bfu_0$,
\item{} it is {\it unique},
\item{} it is {\it holomorphic} in a domain containing $(x_0,\bfu_0,0).$
\end{itemize}

{\it Proof}. See any textbook.
Note that $\bfK$ may be independent of $\varepsilon$,
in which case this is just the existence theorem of Cauchy.

{\it Remark}.
More practically, the canonical form can also be defined as
\begin{equation}
\label{eqODECauchyFormScalar}
{\D^N u \over \D x^N}=K[x,u,u',\dots,u^{(N-1)}].
\end{equation}

{\it Definition}.
Given a differential equation (DE) 
\begin{eqnarray}
& &
\bfE(\bfx,\bfu)=0
\label{eqDEgeneral}
\end{eqnarray}
and a point $\bfu_0$,
the linear DE
\begin{equation}
\label{eqAuxiliary}
\bfE'(\bfx,\bfu^{(0)})\bfv
= \lim_{\lambda \to 0}
 {\bfE(\bfx,\bfu^{(0)} + \lambda \bfv)-\bfE(\bfx,\bfu^{(0)}) \over \lambda}
=0
\end{equation}
in the unknown $\bfv$ is called
the {\it linearized equation} in the neighborhood of $\bfu^{(0)}$
associated to the equation
$\bfE(\bfx,\bfu)=0$.
This was introduced by Darboux \cite{Darboux1883}
under the name ``\'equation auxiliaire''.
The derivative $\bfE'$ is known under various names~:
G\^ateaux derivative,
linearized map, tangent map, Jacobian matrix,
and sometimes Fr\'echet derivative.

Let us define the formal Taylor expansions
\begin{equation}
   \bfu=\sum_{n=0}^{+ \infty} \varepsilon^n \bfu^{(n)},\
   \bfK=\sum_{n=0}^{+ \infty} \varepsilon^n \bfK^{(n)}.
\label{eqPerturbationTaylor}
\end{equation}
The single equation (\ref{eqLemma}) is equivalent to the infinite sequence
\begin{eqnarray}
         n  =  0:\
 {\D \bfu^{(0)} \over \D x}&=&\bfK^{(0)}=\bfK[x,\bfu^{(0)},0]
\label{eqPerturb0}
\\
 n \ge 1:\
 {\D \bfu^{(n)} \over \D x}&=&\bfK^{(n)}=\bfK'[x,\bfu^{(0)},0]
 \bfu^{(n)}
   + \bfR^{(n)}(x,\bfu^{(0)},\dots,\bfu^{(n-1)}).
\label{eqPerturbn}
\end{eqnarray}

At order zero, the equation is nonlinear. 

At order one, the equation,
in the particular important case when $\bfK$ is independent of $\varepsilon$,
is the linearized equation (without rhs, since $\bfR^{(1)}=0$)
canonically associated to the nonlinear equation.
\index{linearized equation}

At higher orders,
this is the same linearized equation with different rhs $\bfR^{(n)}$
arising from the previously computed terms,
and only a particular solution is needed to integrate.
\label{pageTheoremII}

{\it Theorem II}
(Poincar\'e 1890, Painlev\'e 1900, Bureau 1939).
Take the assumptions of previous theorem.
If the general solution of (\ref{eqLemma}) is single valued in $D$ except
maybe at $\varepsilon=0$,
then
\begin{itemize}
\item{} $\varepsilon=0$ is no exception, i.e.~the general
solution is also single valued there,
\item{} every $\bfu^{(n)}$ is single valued.
\end{itemize}

{\it Proof}. See \cite{PaiBSMF} p.~208.
The main difficulty is to prove the convergence of the series.
This theorem remains valid if one replaces
``single valued'' (Painlev\'e version)
by ``periodic'' (Poincar\'e version)
or ``free from movable critical points'' (Bureau version).
\smallskip

This feature (one nonlinear equation (\ref{eqPerturb0}),
one linear equation (\ref{eqPerturbn}) with different rhs)
is a direct consequence of perturbation theory,
it is common to all methods aimed at building necessary stability conditions
(following Bureau, we call {\it stable} an ODE with the PP).
The equations may be differential like (\ref{eqPerturb0})--(\ref{eqPerturbn}),
or simply algebraic.
Moreover, all the methods which we are about to describe
(except the one of Painlev\'e) will reduce the differential
problems to algebraic problems keeping the same feature,
and the overall difficulty will be to solve
{\it one} nonlinear algebraic equation,
then {\it one} linear algebraic equation with a countable number
(practically, a finite number) of rhs.

{\it All the methods} of the Painlev\'e test are applications of the last
theorem~:

\begin{enumerate}
\item
the method of pole-like expansions of Kowalevski and Gambier 
\cite{GambierThese},

\item
the $\alpha-$method of Painlev\'e \cite{PaiBSMF},

\item
the method of Bureau \cite{Bureau1939},

\item
the Fuchsian perturbative method \cite{CFP1993},

\item
the nonFuchsian perturbative method \cite{MC1995}.

\end{enumerate}

These methods establish necessary conditions for the Painlev\'e property 
by building
one or two perturbed equations from the original unperturbed equation,
then by applying the theorem II at a point $x_0$ which is {\it movable}.
This movable point can be either regular (method of Painlev\'e)
or singular noncritical (all the others),
which will require its previous transformation to a regular point
(by a transformation close to $u \to u^{-1}$)
for theorem II to apply.
One is thus led to the equations (\ref{eqPerturbn}),
i.e.~to {\it one linear} DE with a sequence of rhs.
In order to avoid movable critical points in the original equation,
one requires single valuedness in a neighborhood of $x_0$ for~:
the general solution of the linear homogeneous equation,
a particular solution of each of
the successive linear inhomogeneous equations.

\section{Meaning of the negative Fuchs indices}
\indent

Basically, they are just the consequence of a resummation of a series.
In particular, they do {\it not} imply the existence of an essential
singularity.
This is easier to understand if one starts from a given general solution
rather than from a given ODE whose general solution may not be known in
closed form.

The ODE with a meromorphic general solution \cite{CFP1993}
\begin{equation}
E \equiv u'' + 3 u u' + u^3 = 0,\
 u={1 \over x-a} + {1 \over x-b},\ a \hbox{ and } b \hbox{ arbitrary},
\label{eqTwopoles}
\end{equation}
has two families,
\begin{description}
\item (F1) $p=-1,u_0=1$, indices $(-1,1)$, $u =    \chi^{-1} + \dots$,
\item (F2) $p=-1,u_0=2$, indices $(-2,-1)$, $u = 2 \chi^{-1}$,
\end{description}
and the negative index $-2$ must coexist with the meromorphy.
The representation of the general solution (\ref{eqTwopoles})
as a Laurent series of $x-x_0$
is the sum of two copies of an expansion of $1/(x-c)$,
and there exist two expansions of $1/(x-c)$
\begin{eqnarray}
(x-c)^{-1}
{\hskip -3.5 truemm} & = & {\hskip -3.5 truemm}
\sum\limits_{j=- \infty}^{-1} (c-x_0)^{-1-j} (x-x_0)^{j},\
\mod{c-x_0} < \mod{x-x_0}
\\
{\hskip -3.5 truemm} & = & {\hskip -3.5 truemm}
\sum\limits_{j=0}^{+ \infty} - (c-x_0)^{1-j} (x-x_0)^{j},\
\mod{x-x_0} < \mod{c-x_0}.
\end{eqnarray}
The family (F1) corresponds to the sum
(first expansion with $c=a=x_0$) plus (second expansion with $c=b$),
while
the family (F2) corresponds to the sum
(first expansion with $c=a$) plus (second expansion with $c=b$).

The second family series terminates ~: $u=2/(x-x_0)$,
and its algorithmic perturbation into the doubly infinite Laurent series
is handled in Ref.~\cite{CFP1993} by the method of section
\ref{sectionMethodPerturbativeFuchsian}.


\section{The Fuchsian perturbative method}
\label{sectionMethodPerturbativeFuchsian}
\indent

It allows to extract the information contained in the negative
indices \cite{FP1991},
thus building infinitely many necessary conditions
for the absence of movable critical singularities of the logarithmic
type \cite{CFP1993}.

The perturbation which describes it is close to the identity
\begin{equation}
\label{eqPerturbu}
   x \hbox{ unchanged},\
   \bfu= \sum_{n=0}^{+ \infty} \varepsilon^n \bfu^{(n)}:\
   \bfE= \sum_{n=0}^{+ \infty} \varepsilon^n \bfE^{(n)}=0,
\end{equation}
where, like for the $\alpha-$method,
the small parameter $\varepsilon$ is not in the original equation.

Then, the single equation (\ref{eqDEgeneral}) is equivalent to the infinite
sequence
\begin{eqnarray}
\label{eqNL0}
         n  =  0\
\bfE^{(0)}
{\hskip -3.5 truemm} & \equiv & {\hskip -3.5 truemm}
\bfE (x,\bfu^{(0)}) = 0
\\
 \forall n \ge 1\
\bfE^{(n)}
{\hskip -3.5 truemm} & \equiv & {\hskip -3.5 truemm}
\bfE'(x,\bfu^{(0)}) \bfu^{(n)}
             + \bfR^{(n)}(x,\bfu^{(0)},\dots,\bfu^{(n-1)}) = 0,
\label{eqLinn}
\end{eqnarray}
with $\bfR^{(1)}$ identically zero.
From Theorem II, necessary stability conditions are
\begin{description}
\item[-]
the general solution $\bfu^{(0)}$ of (\ref{eqNL0}) has no
movable critical points,
\item[-]
the general solution $\bfu^{(1)}$ of (\ref{eqLinn}) has no
movable critical points,
\item[-]
for every $n\ge 2$ there exists a particular solution of (\ref{eqLinn})
without movable critical points.
\end{description}

Order zero is just the complete equation for the unknown $\bfu^{(0)}$,
so, in order to get some additional information,
one must apply Theorem II for a perturbation different from
(\ref{eqPerturbu}).
One then uses the method of pole-like expansions at this order zero,
{\it only} to obtain the leading term
$\bfu^{(0)} \sim \bfu^{(0)}_0 \chi^{\bfp}$ of all the acceptable families
of movable singularities.

The precise steps of the algorithm are detailed in 
Refs.~\cite{CFP1993,Cargese96Conte}.
The algorithm is purely algebraic, i.e.~one does not perform any
integration.
The expression of each Taylor coefficient $\bfu^{(n)}$ is found to be
a Laurent series of $\chi=x-x_0$,
\begin{eqnarray}
 \bfu^{(n)}
&=& \sum_{j=n \rho}^{+ \infty} \bfu^{(n)}_j \chi^{j+\bfp},\
\label{eqSeriesun}
\end{eqnarray}
in which $\rho$ is the smallest Fuchs index
(therefore lower than or equal to $-1$),
so that the full expression $\bfu$ is a ``full'' Laurent series,
i.e.~one whose powers extend to both infinities.

The Fuchsian perturbative method (as well as the nonFuchsian one which will
be seen section \ref{sectionMethodPerturbativeNonFuchsian})
is useful if and only if
the zeroth order $n=0$ fails to describe the general solution.
This may happen for two reasons.
The most common one is a negative Fuchs index in addition to $-1$ counted
once,
the second, less common one is a multiplicity higher than one for some
family, as in the single family of equation (\ref{eqOrder2WithLog}).

\subsection{The simplest constructive example}
\label{sectionIndex0}
\indent

The equation (\ref{eqOrder2WithLog}) is the simplest example to understand
this method, because

\begin{enumerate}
\item
there exists a movable logarithm,
as shown by the $\alpha-$method (\cite{PaiBSMF} \S 13, p.~221),

\item
the method of pole-like expansions fails to find it,

\item
the Fuchsian perturbative method finds it after a very short computation,
easy to do by hand since all series terminate.

\end{enumerate}

The single family, in the notation of this section, is
\begin{eqnarray}
& &
p=-1,\
E_0^{(0)}= u_0^{(0)} (u_0^{(0)}-1)^2=0,\
\hbox{indices } (-1,0),
\end{eqnarray}
and the order $n=0$ yields the one-parameter series
\begin{eqnarray}
u^{(0)}
& = &
\chi^{-1} \hbox{ (the series terminates)}.
\end{eqnarray}
At order $n=1$,
the derivative of $E$ at point $u=u^{(0)}$ is
\begin{eqnarray}
E'(x,u^{(0)})
& = &
\partial_x^2 + 4 \chi^{-1} \partial_x + 2 \chi^{-2},
\end{eqnarray}
so that the linearized equation for $u^{(1)}$ is of Fuchsian type
(by the way, this is why we call its indices ``Fuchs indices''
and not ``resonances'',
a word which refers to no resonance phenomenon).
The computation of the Laurent series (\ref{eqSeriesun}) for $u^{(1)}$ 
(with $n=1,p=-1,\rho=-1$)
is made in one computer loop
by increasing values of $j$ and exhibits no logarithms.
The result is
\begin{eqnarray}
u^{(1)}
& = &
u_{-1}^{(1)} \chi^{-2} + u_{0}^{(1)} \chi^{-1},\
u_{-1}^{(1)} \hbox{ and }u_{0}^{(1)} \hbox{ arbitrary}.
\end{eqnarray}
The sum $u^{(0)}+ \varepsilon u^{(1)}$ is the beginning of a series which
now depends on two arbitrary parameters,
namely $x_0$ and $\varepsilon u_{0}^{(1)}$
(one can indeed set $u_{-1}^{(1)}$ to zero without loss of generality
since it represents a perturbation of $x_0$).
With the retained value
\begin{eqnarray}
u^{(1)}
& = &
u_{0}^{(1)} \chi^{-1},\
u_{0}^{(1)} \hbox{ arbitrary},\
\end{eqnarray}
the order $n=2$ is
\begin{eqnarray}
E^{(2)}
& = &
E'(x,u^{(0)}) u^{(2)} + 6 u^{(0)} u^{(1)^2} + 4 u^{(1)} u^{(1)'}
\nonumber
\\
& = &
\chi^{-2} (\chi^2 u^{(2)})'' + 2 u_0^{(1)^2} \chi^{-3}
=0,
\end{eqnarray}
and the Laurent series (\ref{eqSeriesun}) for the particular solution
$u^{(2)}$ is found not to exist because of a movable logarithm
\begin{eqnarray}
u^{(2)}
& = &
- 2 u_{0}^{(1)^2} \chi^{-1} (\Log \chi - 1).
\end{eqnarray}
The movable logarithmic branch point is therefore detected in a systematic way
at order $n=2$ and index $i=0$.

The necessity to perform a perturbation arises from the multiple root
of the equation for $u_0^{(0)}$,
responsible for the insufficient number of arbitrary parameters in
the zeroth order series $u^{(0)}$.

\subsection{An example needing order seven to conclude}
\label{sectionOrderSeven}
\indent

In the equation (\ref{eqBureauOrder4}),
the {\it first family} provides, at zeroth order, only a two-parameter
expansion and,
when one checks the existence of the perturbed solution
\begin{equation}
 u=\sum_{n=0}^{+ \infty} \varepsilon^n
 \left[\sum_{j=0}^{+ \infty} u_j^{(n)} \chi^{j-2-3n}\right],
\end{equation}
one finds that coefficients
$u_{20}^{(0)}, u_{-3}^{(1)}, u_{-2}^{(1)}, u_{-1}^{(1)}$
can be chosen arbitrarily,
and, at order $n=7$, one finds two violations \cite{CFP1993}
\begin{equation}
   Q_{-1}^{(7)} \equiv u_{20}^{(0)}   u_{-3}^{(1)^7} = 0,
   Q_{20}^{(7)} \equiv u_{20}^{(0)^2} u_{-3}^{(1)^6} u_{-2}^{(1)} = 0,
\end{equation}
implying the existence of a movable logarithmic branch point.

{\it Remark} \cite{MC1995}.
The value $n=7$ is the root of the linear equation
$n (i_{\rm min}-p) + (i_{\rm max}-p)=-1$,
with $p=-2,i_{\rm min}=-5,i_{\rm max}=18$,
linking the pole order $p$ in the Fuchsian case $c=0$,
the smallest and the greatest Fuchs indices.
It expresses the condition for the first occurrence of a power $\chi^{-1}$,
leading by integration to a logarithm,
in the r.h.s.~$R^{(n)}$ of the linear inhomogeneous equation
(\ref{eqLinn}),
r.h.s.~created by the nonlinear terms $3 u u'' - 4 u'^2 $.

As to the {\it second family}, 
it is useless for the Fuchsian perturbative method,
because the two arbitrary coefficients corresponding to the two Fuchs indices
$(-1,0)$ are already present at zeroth order.

\section{The nonFuchsian perturbative method}
\label{sectionMethodPerturbativeNonFuchsian}
\indent

Every time the family under study has less Fuchs indices
than the differential order $N$,
the Fuchsian perturbation method fails to build a
representation of the general solution,
thus possibly missing some stability conditions.
Such an example is the second family of the equation (\ref{eqBureauOrder4}).
The missing solutions of the auxiliary equation (\ref{eqAuxiliary})
are then nonFuchsian solutions.

Although there is no difficulty to algorithmically compute expansions for
the nonFuchsian solutions,
these are of no immediate help, due to their generic divergence.

There is one situation where some stability conditions can be generated
{\it algorithmically}.
It occurs when the two following conditions are met \cite{MC1995}.

\begin{enumerate}
\item
There exists a particular solution $\bfu=\bfu^{(0)}$
which is known globally, is meromorphic and has at least one
movable pole at a finite distance denoted $x_0$.

\item
The only singular points of the linearized equation $\bfE^{(1)}=0$
are $x=x_0$, nonFuchsian,
and $x=\infty$, Fuchsian.

\end{enumerate}

Then, the property that a fundamental set of solutions $\bfu^{(1)}$
be locally single valued near $\chi=x-x_0=0$
is equivalent to the same property near $\chi=\infty$.
This is the global nature of $\bfu^{(0)}$ which allows the study of the point
$\chi=\infty$, easy to perform with the Fuchsian perturbative method.

An important technical bonus is the lowering of the differential order $N$
of equation $\bfE^{(1)}=0$ by the number $M$ of arbitrary parameters $c$ which
appear in $\bfu^{(0)}$.
Indeed, again since $\bfu^{(0)}$ is closed form,
its partial derivatives $\partial_c \bfu^{(0)}$ are closed form and are
particular solutions of $\bfE^{(1)}=0$,
which allows this lowering of the order.

At each higher perturbation order $n \ge 2$,
one similarly builds particular solutions $\bfu^{(n)}$
as expansions near $\chi=\infty$ and one requires the same properties.

Let us just take one example.
In section \ref{sectionOrderSeven},
the fourth order equation (\ref{eqBureauOrder4}) has been proven to be
unstable after a computation practically intractable without a computer.
Let us now prove this result without computation at all \cite{MC1995}.
For the global two-parameter solution 
\begin{eqnarray}
& &
u^{(0)}=c \chi^{-3} - 60 \chi^{-2},\
(c,x_0) \hbox{ arbitrary},
\end{eqnarray}
the linearized equation
\begin{equation}
E^{(1)} = E'(x,u^{(0)}) u^{(1)} \equiv
 [              \partial_x^4
  + 3 u^{(0)}   \partial_x^2
  - 8 u^{(0)'}  \partial_x
  + 3 u^{(0)''}] u^{(1)} = 0
\label{eqBureauLin}
\end{equation}
has only two singular points $\chi=0$ (nonFuchsian) and $\chi=\infty$
(Fuchsian),
it admits the two global single valued solutions
$\partial_{x_0} u^{(0)}$ and $\partial_c u^{(0)}$,
i.e.~$u^{(1)}=\chi^{-4},\chi^{-3}$.
The lowering by $M=2$ units of the order of the linearized equation
(\ref{eqBureauLin}) is obtained with
\begin{equation}
u^{(1)}=\chi^{-4} v:\
[\partial_x^2 -16 \chi^{-1} \partial_x +3 c \chi^{-3} - 60 \chi^{-2}] v'' = 0,
\end{equation}
and the local study of $\chi=\infty$ is unnecessary,
since one recognizes the confluent hypergeometric equation.
The two other solutions in global form are
\begin{eqnarray}
c \not=0:\
v_1''
& = &
\chi^{-3} {}_{0} F_{1} (24;-3c/\chi)
=
\chi^{17/2} J_{23}(\sqrt{12 c/\chi}),
\\
v_2''
& = &
\chi^{17/2} N_{23}(\sqrt{12 c/\chi}),
\end{eqnarray}
where the hypergeometric function ${}_{0} F_{1} (24;-3c/\chi)$
is single valued and possesses an isolated essential singularity at $\chi=0$,
while the fonction $N_{23}$ of Neumann is multivalued because of a
$\Log \chi$ term.

{\it Remark}.
The local study of (\ref{eqBureauLin}) near $\chi=0$
provides the formal expansions 
for the two nonFuchsian solutions
\begin{equation}
\chi \to 0,\
c \not=0:\
u^{(1)}=e^{\pm \sqrt{-12c/ \chi}}\chi^{31/4} (1 + O(\sqrt{\chi})),
\label{eqBureau4Formal}
\end{equation}
detecting the presence in (\ref{eqBureauLin})
of an essential singularity at $\chi=0$,
but the generically null radius of convergence of the formal series forbids
to conclude to the multivaluedness of $u^{(1)}$.
A nonobvious result is the existence, as seen above,
of a linear combination of the two formal expansions (\ref{eqBureau4Formal})
which is single valued.

An application to the Bianchi IX cosmological model can be found in
Ref.~\cite{LMC1994}.

\section{Miscellaneous perturbations}
\label{sectionMiscellaneous perturbations}
\indent

The differential complexity of the $\alpha-$method explains why
it usually succeeds in case of failure of all the other methods,
which only have an algebraic complexity.

For equation (\ref{eqNoDomOrder3}),
there exists no perturbation satisfying the assumptions of Theorem II
page \pageref{pageTheoremII},
there only exist singular perturbations,
\ie which discard the highest derivative.
Since they however give the correct information, it would be desirable to
extend Theorem II in that direction.
Meanwhile, the reasoning given below cannot be considered as a proof,
and one should refer to the proof of Chazy.

Equation (\ref{eqNoDomOrder3}) is handled by the singular perturbation
\begin{eqnarray}
u &= &
    \varepsilon^{-1} \sum_{n=0}^{+ \infty} \varepsilon^n u^{(n)},\
E = \varepsilon^{-2} \sum_{n=0}^{+ \infty} \varepsilon^n E^{(n)},\
\label{eqPerturmationMisc1}
\end{eqnarray}
which excludes $u'''$ from the simplified equation
and for which all the series happen to terminate,
which makes the computation quite easy
\begin{eqnarray}
E^{(0)} & \equiv & u^{(0)} u^{(0)''} - 2 u^{(0)'^2} =0
\\
u^{(0)} & = & c \chi^{-1},\ \chi=x-x_0,\ (x_0,c) \hbox{ arbitrary},
\\
E^{(1)} & \equiv & c (\chi^{-3} (\chi^2 u^{(1)})'' - 6 \chi^{-4}) = 0,
\\
u^{(1)} & = & 6 \chi^{-1} (\Log \chi-1).
\end{eqnarray}

\section{Conclusion}
\indent

One should not be afraid of negative Fuchs indices.
Indeed, at the level of the linearized equation i.e.~of the term
$n=1$ of a perturbation process,
the sign of these Fuchs indices does not matter at all.
Their presence does not imply the existence of essential singularities,
as shown on an elementary example,
and there do exist algorithmic methods to build no-log conditions
from them,
in the same way as the widely known method of pole-like expansions
builds no-log conditions from the positive indices.

\section*{Acknowledgments}
\indent

The author would like to thank 
the Bharatidasan University for its warm hospitality and support,
and the Minist\`ere des affaires \'etrang\`eres for travel support.

\end{document}